\documentclass[fleqn,10pt]{wlscirep}

\newcommand{\degC}[1]{$#1~^{\circ}\text{C}$}
\newcommand{\mn}[1]{#1~mN~m$^{-1}$}
\newcommand{\mm}[1]{#1~mm~s$^{-1}$}

\usepackage{microtype}

\title{Interfacial mechanisms for stability of surfactant-laden films}

\author[1,*]{M. Saad Bhamla}
\author[1]{Chew Chai}
\author[2]{Marco A. \`{A}lvarez-Valenzuela}
\author[3]{Javier Tajuelo}
\author[1]{Gerald G. Fuller}
\affil[1]{Stanford University, Department of Chemical Engineering, Stanford, 94305, USA}
\affil[2]{Universidad Carlos III de Madrid, Department of Mechanical Engineering, Leganes,  28911 , Spain}

\affil[3]{Universidad Nacional de Educaci{\'o}n a Distancia (UNED),Departamento de F{\'i}sica Fundamental, Madrid,  28040 , Spain}

\affil[*]{bhamla@stanford.edu}

\begin{abstract}
Thin liquid films are central to everyday life. They are ubiquitous in modern technology (pharmaceuticals, coatings), consumer products (foams, emulsions) and also serve vital biological functions (tear film of the eye, pulmonary surfactants in the lung). A common feature in all these examples is the presence of surface-active molecules at the air-liquid interface. Though they form only molecularly-thin layers, these surfactants produce complex surface stresses on the free surface, which have important consequences for the dynamics and stability of the underlying thin liquid film. Here we conduct simple thinning experiments to explore the fundamental mechanisms that allow the surfactant molecules to slow the gravity-driven drainage of the underlying film. We present a simple model that works for both soluble and insoluble surfactant systems. We show that surfactants with finite surface rheology influence bulk flow through viscoelastic interfacial stresses, while surfactants with inviscid surfaces achieve stability through opposing surface-tension induced Marangoni flows.

\end{abstract}
\begin{document}

\flushbottom
\maketitle
%
%
\thispagestyle{empty}

\section*{Introduction}

Stability and drainage of thin surfactant films is relevant across various disciplines: industrial applications including engineered foams and emulsions\cite{Behera:2014cj}, fundamental physics of bubbles\cite{Debregeas:1998gq,Feng:2014hu,Bird:2010dm}, bio-foams in aquatic animal nests\cite{Cooper:2010bd}, and physiological systems including the human tear film\cite{2014BhamlaSoftMatter} and pulmonary surfactants\cite{2015SMat...11.8048H}. However, the drainage rate of these thin films depends critically on the mechanism through which these films are stabilized, which in turn is strongly coupled to the chemical composition of the surfactants. 

The majority of past literature has looked at the stability of thin films in presence of soluble amphiphiles, including drainage from horizontal films\cite{Radoev:1983eb,Nikolov:1990fx}, drainage of vertical films based on Frankel's law\cite{Mysels:1962ba,Lyklema:1965bi} and film stability in fiber coating experiments\cite{Shen:2002dd}. In comparison, the problem of drainage in presence of insoluble surfactants has been studied relatively less due to experimental challenges; the majority of investigations by Naire and coworkers  focused on mathematical models to study the drainage of vertical thin films in the presence of insoluble surfactants\cite{Naire:2000gn,Naire:2001gx,Naire:2004jc}. Past work by Joye et al. also presents numerical simulations and linear stability analysis to explore the role of surface rheological parameters\cite{Joye:1994di,Joye:1996kw}. However, there is a need for a simple experimental platform that can systematically compare both soluble and insoluble surfactants, with varying surface rheologies and quantify the drainage dynamics using a simple theoretical model.

\begin{figure}[ht]
\centering
\includegraphics[width=.4\linewidth]{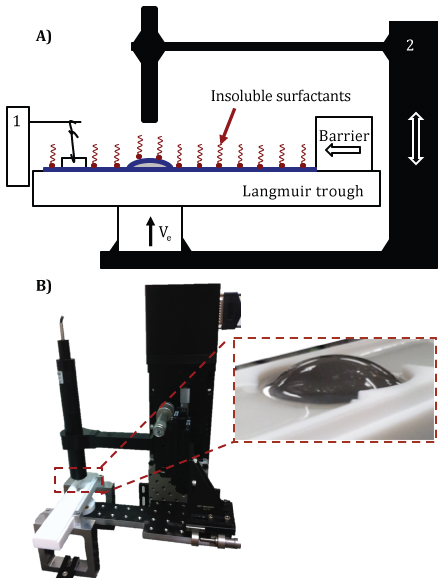}
\caption{\textbf{Experimental platform} Photograph (A) and schematic (B) of the drainage platform. For the insoluble surfactant experiments, the glass dome is initially submerged in the PBS-filled Langmuir trough (white, teflon container) and DPPC is spread at the air-liquid interface. DPPC is then compressed to the desired surface pressure using a single Delrin barrier and the surface pressure is monitored using a paper Wilhelmy balance (1). For the soluble surfactant experiments, the Langmuir trough is filled with SDS solution of desired concentration. In both cases, the measurement commences once the glass dome is elevated through air-liquid interface with a computer controlled motorized stage (2). A high speed interferometer (black tube) captures the thickness of the draining films as a function of time at the apex of glass dome.}
\label{Fig:iDDrOP} 
\end{figure}

Here we utilize a simple setup (Fig.~\ref{Fig:iDDrOP}) to measure the drainage dynamics of surfactant-laden aqueous films. Thin films of liquid are created by elevating an initially submerged curved glass substrate through the air-liquid interface at controlled velocities ($V_{e}$). A high-speed interferometer enables measurement of the varying film thickness at the apex of the film. We employ two heavily-studied commercial surfactants: 1,2-dipalmitoyl phosphatidylcholine (DPPC), an insoluble surfactant that forms viscoelastic interfaces\cite{Choi:2011kz,2014SMat...10..175H} and sodium~dodecyl~sulfate (SDS), a soluble surfactant that forms inviscid interfaces\cite{Zell2014}. We also show by surface flow visualization that the viscoelasticity of DPPC films resists surface deformation and creates an \textit{immobile} interface at high surface pressures, while the SDS films are more fluid-like and yield extremely \text{mobile} interfaces. The remarkably different surface properties between DPPC and SDS allow us to systematically explore the role of surface mobility on drainage dynamics.

\section*{Results}

\subsection*{Theoretical hydrodynamic model for draining films with complex interfaces}\label{SS:theory}
Consider a hemispherical glass dome that is raised through a bulk of liquid that results in the capture of a thinning liquid film with a complex surfactant-laden interface. Deformation of the interface leads to interfacial stresses that need to be accounted for in the hydrodynamic model capturing the evolution of this thin film. Analytical solution to this thin-film draining problem is well-established in our past work\cite{2014BhamlaSoftMatter,2015SMat...11.8048H}. From this mathematical analysis, the following single-parameter expression predicts the dimensionless thickness $H$ of the film at the apex, due to gravity, as a function of time:
\begin{equation}
H = \frac{h}{h_{0}}=\frac{1}{\sqrt{1+4\alpha\tau}},
\label{Fuller}
\end{equation}
\noindent where $\tau=t\rho g h_{0}^2/(\eta R)$ is the dimensionless time normalized by known experimental variables: bulk viscosity ($\eta$), substrate curvature ($R$), density ($\rho$), gravitational constant ($g$), time ($t$) and initial film thickness ($h_{0}$). The parameter $\alpha$ characterizes the fluidity of the interface and is directly coupled to the total Boussinesq number, $Bq=Bq_{s}+Bq_{d}+Ma$, where $Bq_{s}$, $Bq_{d}$ and $Ma$ are dimensionless numbers capturing the influence of surface shear viscosity, surface dilatational viscosity and Marangoni stresses. These numbers are defined as follows: $Bq_{s}=\eta_{s}/(\eta h_{0})$, $Bq_{d}=\eta_{d}/(\eta h_{0})$ and $Ma=R^{2}/(\rho g h_0^{3})  \nabla{\sigma}$, where $\eta_{s}, \eta_{d}$ are the surface shear and dilatational viscosity due to the surfactant-molecules, $\nabla{\sigma}$ is the gradient in the surface-tension across the air-liquid interface\cite{2015SMat...11.8048H}. 

The total Boussinesq number captures the relative drag from the interface to the bulk due to combined influence of shear, dilatational and Marangoni stresses. Asymptotic limits of this model indicate that for a surfactant-free interface $\alpha=0.33$ (clean water: $Bq=0$), while for an extreme viscous surfactant-laden interface $\alpha = 0.08$ ($Bq\rightarrow\infty$)\cite{2014BhamlaSoftMatter}. The model in Eq.~\ref{Fuller} has been shown to work well with commercial lung surfactant systems that contained a mixture of unknown soluble and insoluble components\cite{2015SMat...11.8048H}. The purpose of this paper is to employ commonly-used systems that can be systematically controlled to provide conclusive evidence of stabilizing interfacial mechanisms.  

The parameter $\alpha$, thus allows us to evaluate the role of interfacial contributions, including surface viscosities ($Bq_{s},Bq_{d}$) and surface-tension gradients ($Ma$), on the draining behavior of thin films. However, by itself, $\alpha$ is insufficient to inform the relative contributions of surface rheology versus Marangoni effects. Thus, we further conduct surface flow visualization experiments that help in distinguishing between these different interfacial phenomena.

\subsection*{Insoluble surfactant: DPPC}

\begin{figure}[htbp]
\centering
\includegraphics[width=.9\textwidth]{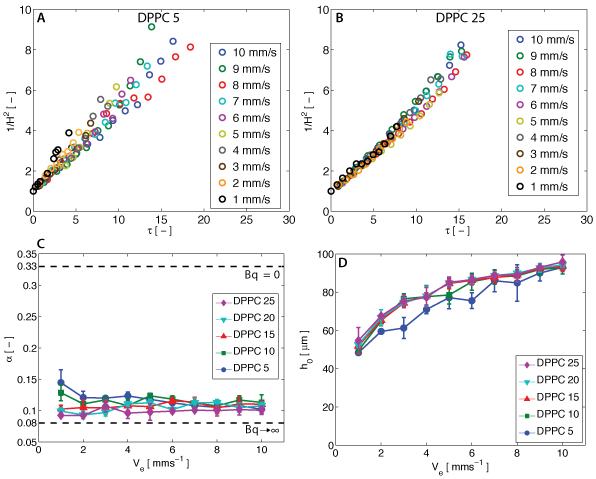}
\caption{\label{F:DPPCsummary} \textbf{DPPC drainage experiments}. [A,B]  Dimensionless variable ($ 1/H^{2}= (h_{0}/h)^{2})$ as a function of rescaled time ($\tau$) for DPPC at \mn{5}  and \mn{25} at various elevated velocity ($V_{e}$) ranging from \mm{1-10}. [C] Summary of the value for the fitting parameter $\alpha$ of DPPC at various surface pressures.  [D] Summary of the initial height capture of the aqueous film laden with DPPC at different surface pressures. The standard deviation is calculated from three independent trials.}
\label{Fig:DPPC}
\end{figure}

For insoluble surfactants, it is common to connect the surface concentration through the surface pressure $\Pi=\gamma_{0} - \gamma$, where $\gamma$ and $\gamma_{0}$ are the surface tension with and without surfactants, respectively. For DPPC films, increasing the surface pressure has the effect of increasing both the surface shear and dilatational moduli\cite{BhamlaJCIS2015,2014SMat...10..175H,Anton:2013fl,Kim:2013fs}. We conduct a surface pressure sweep from $\Pi=5-$\mn{25}, and present the data for $\Pi=5$ and \mn{25} in Fig.~\ref{Fig:DPPC}A,B. In these figures, the inverse-squared film thickness $1/H^{2}= (h_{0}/h)^{2}$ is plotted as a function of $\tau$, for different films raised at varying elevation velocities, $V_{e}=1-$\mm{10}. At both surface pressures, $\Pi=5$ and \mn{25}, the data trends linearly, yielding extremely good fits to our simple drainage model. Using Eq.~\ref{Fuller}, the corresponding $\alpha$ values are shown in Fig.~\ref{Fig:DPPC}C, for the span of surface pressures examined.

At higher surface pressures ($\Pi>$\mn{10}), $\alpha=0.1$ is in good agreement to the theoretical prediction for a no-slip interfacial boundary condition ($\alpha=0.08, Bq\rightarrow\infty$) and expected for DPPC due to its finite surface viscoelastic properties ($Bq_{s},Bq_{d}>>Ma$)\cite{BhamlaJCIS2015,Anton:2013fl}. In the absence of surfactants, the films should drain rapidly ($Bq=0$) and have no added interfacial stabilization mechanism. We observe this for pure water films as they drain almost instantly ($<50$~ms), and cannot be captured using our interferometric technique. It is important to note that for DPPC, the dilatational viscosity is three orders of magnitude larger than its shear counterpart, and thus may play a more dominant role in stabilization ($Bq_{d}>Bq_{s}$)\cite{Anton:2013fl}. However, at lower surface pressures  ($\Pi<$\mn{10}), DPPC has a more fluid-like interface, and obtaining the same value of $\alpha=0.1$ hints that perhaps Marangoni stresses play a significant role in stabilization. Moreover, the  $\alpha$ values also show slightly larger variations ($0.09-0.15$) at these low surface pressures. We will later demonstrate that Marangoni effects become important for DPPC films at low surface pressures ($Ma>Bq_{s},Bq_{d}$), using surface flow visualization experiments.

We also report the initial height $h_{0}$ of the DPPC films as a function of elevation velocity ($V_e$) in Fig.~\ref{Fig:DPPC}D. We find that increasing the elevation speed results in the capture of thicker DPPC films. This is expected due to the increased lubrication pressure in the thin film ($h\sim R \sqrt{Ca}$), where $Ca$ is the Capillary number defined as $Ca=\eta V_{e}/\gamma$~\cite{Leal:2007tf}. Thus, the faster the elevation velocity, the thicker the film that is captured. 

\subsection*{Soluble surfactant (SDS)}
\begin{figure}[htbp]
\centering
\includegraphics[width=.9\textwidth]{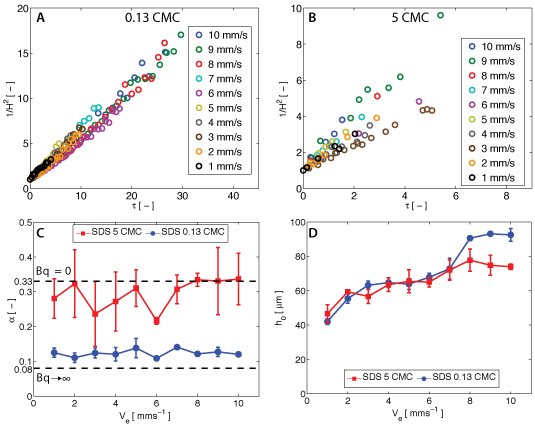}
\caption{\textbf{SDS drainage experiments}. [A,B] Dimensionless variable ($ 1/H^{2}= (h_{0}/h)^{2})$ as a function of rescaled time ($\tau$) for SDS at 0.13~cmc and 5~cmc at various elevated velocity ($V_{e}$) ranging from \mm{1-10}. [C] Summary of the value for the fitting parameter  $\alpha$ of SDS at 0.13~cmc and 5~cmc.  [D] Summary of the initial height capture of SDS film at 0.13~cmc and 5~cmc. The standard deviation is calculated from two independent trials.}
\label{Fig:SDS}
\end{figure}

For SDS films, elevation velocity sweeps ($V_{e}=1-$\mm{10}) are conducted for two concentrations, below and above the cmc. We again plot the inverse-squared film thickness $1/H^{2}$ as a function of $\tau$ for films elevated at different $V_{e}$ in Fig.~\ref{Fig:SDS}(A,B). For both cases, above and below CMC, the data again exhibits linear trends indicating  that our simple model also fits well to the soluble surfactants systems. The corresponding $\alpha$ values for SDS are shown in Fig.~\ref{Fig:SDS}C.

For SDS below its CMC, we obtain $\alpha \sim0.10$ which is similar to $\alpha$'s obtained for DPPC films. Thus, SDS and DPPC both stabilize against film drainage. However, it is well-established that SDS films have inviscid surfaces\cite{Zell2014}, ruling out the role of surface rheology ($Bq_{s},Bq_{d}=0$). Thus, this stabilization has to occur via strong surface-tension gradients or Marangoni flows ($Ma>0$), which will indeed be confirmed using surface flow visualization in the next section.

Above the cmc, SDS films are less stable and drain rapidly with $\alpha\sim0.3$. These drainage rates have significant error bars as the films drain almost instantly and only a few data-points can be recorded - the duration of drainage is $<2$~s (see SI, Fig1). This faster drainage of SDS films at concentrations above the cmc is due to a weakened Marangoni stress resulting from increased repulsions of ionic micelles and reduced surface diffusion, and is well-studied for SDS\cite{Rao:1982gu,Sett:2013kf,Karakashev:2010dr,Berg:2005hg,Patist:2001fs}. 

Finally, we also show the initial entrained height ($h_0$) of the SDS films as a function of $V_{e}$ in Fig. \ref{Fig:SDS}D. At both concentrations, faster elevation velocities lead to larger initial thicknesses, similar to our observation for DPPC films, due to increased lubrication pressures. We also observe that for $V_{e}>6$~mm~s$^{-1}$, films above the cmc concentration are relatively thinner than solutions below the cmc. Similar decreases in thickness have been observed in SDS entrainment experiments on fibers and are a consequence of micellar kinetics that reduces the Marangoni-induced stress, resulting in thinner films\cite{Shen:2002dd,Stebe:1991cr,Patist:2001fs}.

\subsection*{Surface flow visualization}\label{SS:Marangoni}

\begin{figure}[htbp]
\centering
\includegraphics[width=0.3\columnwidth]{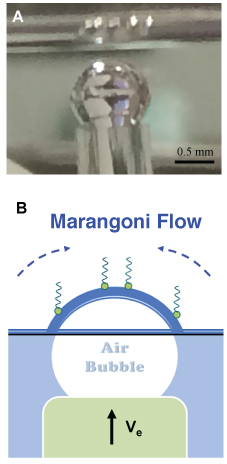}
\caption{\textbf{Surface flow visualization} Photograph (A) and schematic (B) of the surface visualization setup. Instead of a glass substrate, an air bubble is elevated through air-water interface. Thin film color interference patterns are clearly visible under diffused white-light illumination, due to enhanced refractive index mis-match.}
\label{Fig:iDOME}
\end{figure}

\begin{figure*}[htbp]
\centering
\includegraphics[width=.8\textwidth]{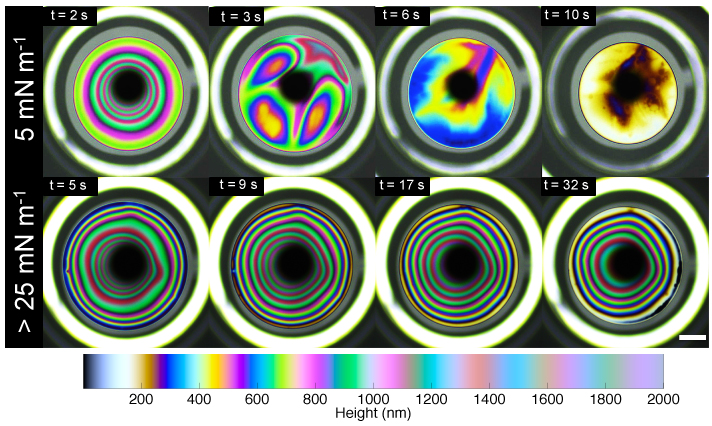}
\caption{\textbf{DPPC surface visualization} Images of the color interference patterns captured for DPPC at two different surface pressures, $\Pi=$\mn{5} and $>$\mn{25} using the surface flow visualization setup. The colormap is a visual tool to determine the corresponding thickness of individual vibrant color. The dark black spot in the center of each frame is the reflection of the camera and the white bright ring at the periphery is the edge of the glass capillary. The scale bar shown is 0.25~mm.}
\label{Fig:DPPC_bubble}
\end{figure*}

\begin{figure*}[htbp]
\centering
\includegraphics[width=.8\textwidth]{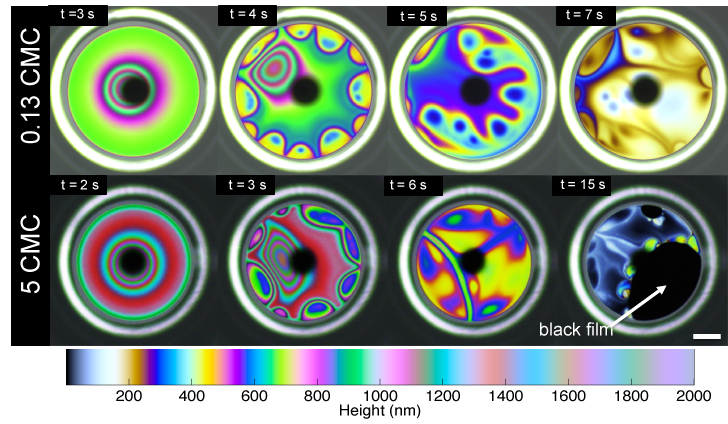}
\caption{\textbf{SDS surface visualization} Snapshots of interference patterns observed for SDS at 0.13 cmc and 5 cmc. The images are attained using the surface flow visualization setup. The colormap is a guide to relate individual vibrant color to its corresponding thickness. The dark black spot in the center of each frame is the reflection of the camera and the white bright ring at the periphery is the edge of the glass capillary. The scale bar shown is 0.25~mm.}
\label{Fig:SDS_bubble}
\end{figure*}

In order to access the dominating influences of surface rheology ($Bq_{s},Bq_{d}$) vs. Marangoni stresses ($Ma$) for DPPC and SDS films, we substitute the glass dome by an air bubble (Fig.~\ref{Fig:iDOME}). The air bubble provides an enhanced refractive index contrast, resulting in extremely vibrant thin-film color interference patterns under white-light illumination. These color fringes vary in space and time, revealing the surface flows or lack thereof, and can be seen clearly in the attached Supplementary Information Movies. Compiled time snap-shots of DPPC and SDS are shown in Figs.~\ref{Fig:DPPC_bubble} and \ref{Fig:SDS_bubble}, respectively. It is useful to use the previously defined terminology for similar surface flows observed using a Scheludko-Exerowa setup by Joye et al.~\cite{Joye:1994di} These authors categorize the patterns as: \textit{symmetric}, associated with with surfaces possessing large surface shear and dilatational viscoelasticities; and \textit{asymmetric}, associated with surfaces with low surface rheology, that yield to surface-tension induced Marangoni flows.

\begin{figure*}[htbp]
\centering
\includegraphics[width=0.7\textwidth]{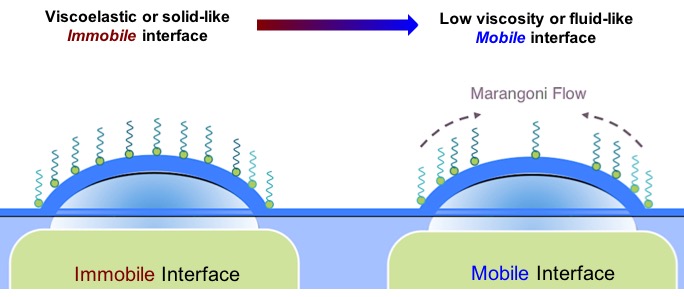}
\caption{\textbf{Surfactant stability mechanisms} Schematic summarizing the two different stabilizing interfacial mechanisms for surfactant films: Viscoelastic interfaces create \textit{immobile} films that reduce drainage through surface stress dissipation, while surface inviscid surfaces create \textit{mobile} interfaces and create surface-tension induced Marangoni flows that counter the bulk-flow direction. }
\label{Fig:Marangonischematic}
\end{figure*}

For DPPC, both the symmetric and asymmetric drainage patterns can be observed, as its surface rheology is a strong function of surface pressure. At low surface pressures ($\Pi$=\mn{5}), the DPPC film is initially symmetric, with circular fringe patterns ($t<2$~s). However, this stable pattern is quickly deformed by plumes of surfactant rising from the periphery. These surface flows are fundamentally similar to rising plumes in vertical soap films, commonly referred to as `marginal regeneration'~\cite{Nierstrasz:2001fw,Nierstrasz:1999dq,Nierstrasz:1998exa,Stein:1991gda,Hudales:1990dg}. These plumes are driven by surface-tension gradients and can be explained as follows. The dilation of the air-liquid interface creates new surface area, resulting in a lower density of the surfactant near the apex. This creates a local area of high surface tension which pulls liquid from the bulk liquid (periphery) that is at a lower surface tension as illustrated in Fig.~\ref{Fig:Marangonischematic}. This results in rising plumes from the periphery towards the apex that re-distribute the DPPC molecules, resulting in a heterogenous pattern that slowly thins, and ultimately the film bursts at $t\sim10$~s. 

 At higher surface pressures $\Pi>$\mn{25}, DPPC films exhibit significant surface viscoelasticity, which suppresses all Marangoni flows. We thus observe stable and symmetric patterns that persist over the entire duration of drainage $t=30$~s, until the film bursts. Similar stable patterns have also been observed for other viscoelastic lipids\cite{2015SMat...11.8048H}. This reinforces our observation that for DPPC films, increasing the surface pressure transitions the stabilization mode from a Marangoni-dominated at low pressures to viscoelasticity-dominated at higher surface pressures.

SDS films at 0.13~cmc and 5~cmc do not possess any measurable surface rheology\cite{Zell2014}. Thus, we would expect these films to exhibit asymmetric drainage, dominated by Marangoni flows. Indeed, similar to DPPC at low surface pressures, we observe rising plumes from the periphery towards the apex driven by surface-tension gradients that ultimately lead to stabilization of the draining bulk film (see Fig.~\ref{Fig:SDS_bubble}). Moreover, for both concentrations, the films initially show the formation of an unstable dimple in the center, which rapidly gets sucked towards the periphery of the film. This `fleeting dimple' has been widely studied in the past literature\cite{Chan:2011vu,Joye:1992jm,Platikanov:1964ek,Hartland:1977ik}. For the solutions above the cmc, we also observe formation of black films before rupture. Thus, as quantitatively described in our drainage experiments, SDS films above and below cmc stabilize via Marangoni stresses, as the surfaces do not posses any surface viscoelasticity to quench these flows.

\section*{Discussion}
It is common knowledge that the presence of surfactants (e.g. soap) extend the life span of thin films (e.g. bubbles). However, a systematic comparison between the drainage of soluble and insoluble surfactants has not been previously presented. We offer an experimental platform to measure the gravity-driven drainage dynamics of DPPC (insoluble) and SDS (soluble) films, as well as a simple model that can be employed to evaluate the influence of interfacial phenomena. We show that the presence of both DPPC and SDS at the air-liquid interface increases the stability of thin films. Specifically, DPPC films are stabilized through interfacial rheology at high surface pressures, resulting in immobile surfaces and Marangoni stresses at low surface pressures, resulting in mobile surfaces. Thus, the surface pressure of DPPC serves as a control for switching surface mobility on and off. Finally, SDS films are stabilized purely through Marangoni effects, resulting in mobile surfaces both above and below CMC. We thus show that soluble and insoluble surfactant systems exploit two fundamentally unique interfacial mechanisms to achieve the same result: thin film stability.

\section*{Methods}

\subsection*{Surfactants}

Two commercially available surfactants are compared: 1,2-dipalmitoyl phosphatidylcholine (DPPC) and sodium dodecyl sulfate (SDS). DPPC is purchased from Avanti Polar Lipids Inc. (Alabaster, AL) in 25~mg~mL$^{-1}$ glass vials. We diluted it to a concentration of 1~mg~mL$^{-1}$ in chloroform (Sigma-Aldrich, St. Louis, MO) and stock solutions were kept in freezer until use. To achieve a desired surface pressure, we spread DPPC at the interface, and compressed using a teflon barrier. SDS (Sigma-Aldrich, St. Louis, MO) solutions were prepared using phosphate buffer saline (PBS, 50 mM, pH 7.0; Gibco) to a desired concentrations of 0.28 mM and 10 mM. These particular concentrations were chosen to investigate the effect of micelles on the stability of draining films. It is important to note that the critical micelle concentration (cmc) of SDS in buffered solution is 2~mM\cite{Sureshbabu:2008cy}.

\subsection*{Experimental Setup}
 A photograph and a schematic of the apparatus used to characterize the drainage of thin films are shown in Fig. \ref{Fig:iDDrOP}, which is similar to the drainage apparatus used previously for study of lung surfactants\cite{2015SMat...11.8048H}, and a slight modification for the study of tear film\cite{2014BhamlaSoftMatter}. Unlike the titanium dome with contact lens, the solid curved glass dome (Newport KPX579, with a curvature of 19.9 mm) is mounted on a pedestal. It is initially submerged in the solution filled Teflon mini-Langmuir trough that is fixed onto a stationary support structure. The trough enables the spreading of insoluble surfactant (DPPC) on aqueous subphase at a controlled surface pressure.  Having a controlled surface pressure is important as the interfacial shear rheology is a strong function of surface pressure. It is also important to mention  that the reported experiments were all conducted at room temperature (\degC{23}). In the case of DPPC, the surface pressure is continuously monitored using a paper Wilhelmy balance connected to a surface pressure sensor (KSV NIMA Ltd., Helsinki, Finland), and only a small deviation of $ \pm$\mn{0.3} is observed. However, for SDS, the glass dome is initially submerged in SDS-filled trough, and the adsorption equilibrium is awaited.  Once the desired surface pressure is reached in both cases, the dome is elevated using a motorized stage through the interface, and captures a thin liquid film. The thickness of this film is measured using a high speed white light interferometer (F70, Filmetrics, USA) combined with a halogen light (Fiber-Lite PL-800). 

The following protocol was followed while conducting the experiment. The glass dome is initially positioned approximately 130~$\mu$m below the interface. The dome is then raised 2.5 mm  at various speed, $V_{e}$, ranging from \mm{1-10} and the thickness of the film, $h$, is captured by the interferometer. For every dataset, the height versus time data was fitted with the theoretical model to attain a characteristic value of the fitting parameter $\alpha$.

\subsection*{Visualization of surface flows}

To assess and visualize the interfacial and Marangoni stress-induced flows, the drainage setup was slightly modified. The elevating glass dome was replaced by an elevating air~bubble (1.1~mm, dia) as shown in Fig. \ref{Fig:iDOME}. The air bubble, generated at the tip of the glass capillary (Drummond Micropipette, Fisher Scientific Inc., MA, USA) approaches an air-liquid interface in the presence of surfactants. The use of an air bubble instead of a glass dome provides a better index of refraction contrast. The air bubble is initially positioned approximately 100~$\mu$m below the interface. The air bubble is then elevated at the speed of \mm{0.3} by a vertical distance of 0.6 mm for DPPC and 1 mm for SDS. These particular vertical distances are chosen to ensure that only a fraction of the bubble cap is exposed at the interface. Under white light illumination (420-780~nm), the color interference patterns of these thin curved films ($< 1~\mu$m) are captured using a color CCD camera.

\bibliography{Drainage_new2}

\section*{Acknowledgements}

The authors would like to thank Daniele Tammaro and John Frostad for useful discussions. J. T. acknowledges a grant from UNED's Researchers Formation Program and partial support from MINECO (Grant FIS2013-47350-C5-5-R).

\section*{Author contributions statement}
MSB and GGF conceived and MSB, CC, MAV constructed the experimental apparatus. CC, MAV and JT performed the experiments. MSB and CC analyzed the data. MSB, CC and GGF wrote the manuscript. All authors reviewed the manuscript. 

\section*{Additional information}
 \textbf{Competing financial interests} The authors declare no competing financial interests. 



\end{document}